\documentclass[10pt,conference]{IEEEtran}

\usepackage{cite}
\usepackage{amsmath,amssymb,amsfonts}
\usepackage{algorithmic}
\usepackage{graphicx}
\usepackage{textcomp}
\usepackage{url}
\usepackage{xcolor}
\def\BibTeX{{\rm B\kern-.05em{\sc i\kern-.025em b}\kern-.08em
    T\kern-.1667em\lower.7ex\hbox{E}\kern-.125emX}}

\usepackage[inline]{enumitem}
\usepackage{makecell}
\usepackage{amsthm}
\usepackage{stmaryrd}
\usepackage{algorithm}
\usepackage{wrapfig}
\usepackage{mathpartir}
\usepackage{booktabs}
\usepackage{multirow}
\usepackage{subcaption}
\usepackage{pifont}
\usepackage{tcolorbox}
\tcbset{
  rqbox/.style={
    colback=gray!10,
    colframe=black,
    boxrule=0.5pt,
    arc=2pt,
    left=0pt, right=0pt, top=3pt, bottom=3pt
  }
}
\usepackage[switch]{lineno}

\definecolor{figblue}{RGB}{72, 116, 202}

\theoremstyle{definition}
\newtheorem{definition}{Definition}[section]
\newcommand{\head}[1]{\noindent\textbf{#1}}
\newcommand{\op}[1]{\textsf{\small#1}}
\newcommand{\semlabel}[1]{{\normalfont\small\texttt{#1}}}

\usepackage[hidelinks,bookmarks=true]{hyperref}

\begin{document}

\title{Programming-by-Example for Batch-Editing Collision Meshes in 3D Software}

\author{%
\IEEEauthorblockN{
Gengyang Xu\textsuperscript{1},
Dongwei Xiao\textsuperscript{1,\dag},
Hengcheng Zhu\textsuperscript{1,\dag},
Yiteng Peng\textsuperscript{1}\\
Wei Meng\textsuperscript{2},
Shuai Wang\textsuperscript{1},
Shing-Chi Cheung\textsuperscript{1}
}
\IEEEauthorblockA{
\textsuperscript{1}The Hong Kong University of Science and Technology
\quad
\textsuperscript{2}The Chinese University of Hong Kong\\
\{gxuah, dxiaoad, ypengbp, shuaiw, scc\}@cse.ust.hk,
hzhuaq@connect.ust.hk,
wei@cse.cuhk.edu.hk\\
\textsuperscript{\dag}Corresponding authors.
}
}

\maketitle

\begin{abstract}
	As 3D software proliferates, software artifacts now extend beyond code and 2D user interfaces to include 3D assets.
	Among these assets, collision meshes are critical as they define the geometry used by physics engines for collision detection and physical interaction.
	Although existing tools can automatically generate collision meshes from visual meshes, they often fail to capture the intended interaction behavior.
	As a result, developers need to manually edit many heterogeneous collision meshes, a process that is time-consuming and challenging to scale.

	To address this problem, we present a neuro-symbolic program synthesis approach for batch-editing collision meshes.
	We formulate the task as a programming-by-example problem: given a family of collision meshes with the same editing intent and a small number of user demonstrations, our approach synthesizes a reusable program that captures the editing intent and applies it to non-demonstration meshes.
	We implement this in a tool named MeshForge, and evaluate it across 24 tasks on 600 collision meshes.
	MeshForge successfully synthesizes 23/24 tasks, requiring 2.2 demonstrations and 3.5 seconds of synthesis time on average.
\end{abstract}

\begin{IEEEkeywords}
	3D software, vision-language model, collision, domain-specific language, programming-by-example.
\end{IEEEkeywords}

\section{Introduction}
\label{sec:introduction}
\begin{figure}[t]
	\centering
	\includegraphics[width=0.95\linewidth]{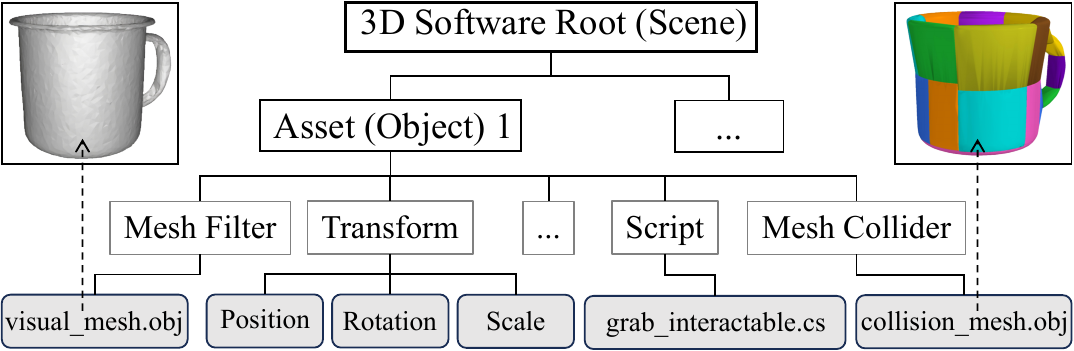}
	\caption{Typical hierarchical structure of 3D software.
		An asset consists of a visual mesh and a collision mesh, the latter consisting of convex hulls (each colored block = one hull).}
	\label{fig:3d-software-structure}
\end{figure}

Driven by advances in graphics hardware and physics engines, 3D software has proliferated across domains such as robotics, embodied AI, digital twins, and VR/AR, exposing new software-engineering challenges in the creation, maintenance, and testing of 3D software~\cite{li2026,li2024less,wang2022vrtest}.
One such challenge concerns the collision behavior of 3D assets.
As shown in Fig.~\ref{fig:3d-software-structure}, a 3D asset commonly has both a \textit{visual mesh} for rendering and a \textit{collision mesh}\footnote{A collision mesh is a simplified mesh to reduce computational overhead of physics simulation, typically represented as convex hulls (hereafter ``hulls'').} for collision detection and physical interaction.
While visual meshes are often created by 3D artists through modeling or scanning, collision meshes are typically derived from visual meshes using geometry-driven convex decomposition algorithms~\cite{unreal_engine_collision_setup,mamou2016volumetric,wei2022approximate}.
However, geometric approximation alone is often insufficient for the intended interaction, so the generated collision meshes frequently require developer post-editing.
Fig.~\ref{fig:introExamples} illustrates two such needs: hulls inside a drawer cavity can block object placement even when surrounding geometry is well approximated, and a generated drawer-handle proxy may be unsuitable for robotic grasping.
Such limitations are also noted in prior work~\cite{andrews2024navigation,wei2022approximate}.

One might expect that increasing decomposition fidelity would address these issues. However, this is not a reliable fix: higher fidelity may improve the problematic region but over-fragment already adequate regions.
In practice, developers typically fix these issues by manually editing relevant regions of generated collision meshes~\cite{li2025evaluating,Conkey2021}.
Such per-mesh editing is labor-intensive and redundant: assets in the same category may share the same \emph{editing intent} despite geometric variations.
This calls for \emph{batch-editing programs} that generalize the editing intent across different assets within the same category.
While current mesh editing tools (e.g., Blender API~\cite{blender_api}) support certain forms of whole-mesh batch-editing (e.g., scaling, rotation, and format conversion), they cannot identify task-relevant hulls across geometrically heterogeneous meshes.

Building on the success of program synthesis~\cite{li2022push,jain2022jigsaw,wu2025question,pan2021can} and multimodal models~\cite{chen2020unblind,cooper2021takes,huang2025seeing,yan2024semantic,zhang2025ufo,lu2025axis} in software engineering (SE), we present MeshForge, a neuro-symbolic program synthesis approach for collision mesh batch-editing.
MeshForge adopts a programming-by-example (PBE) formulation: given a family of collision meshes sharing the same editing intent, users demonstrate the desired edits on a few representative meshes through a graphical user interface (GUI), and MeshForge synthesizes a reusable program that applies it to other meshes in the same asset category.
Specifically, MeshForge employs a pre-trained multimodal model for perception and a symbolic component to generalize editing logic.
MeshForge models a batch-editing program as one or more \texttt{\small extractor-action} pairs: extractors identify target hulls, and actions apply the corresponding edits.
During demonstration, MeshForge records the user-applied actions and synthesizes the extractor logic to generalize across the batch.
Although PBE has been successfully applied to domains such as data extraction, mobile app synthesis, and program transformation~\cite{le2014flashextract,meng2013lase,rolim2017learning}, this formulation exposes three challenges:

\begin{figure}[t]
	\centering
	\includegraphics[width=\linewidth]{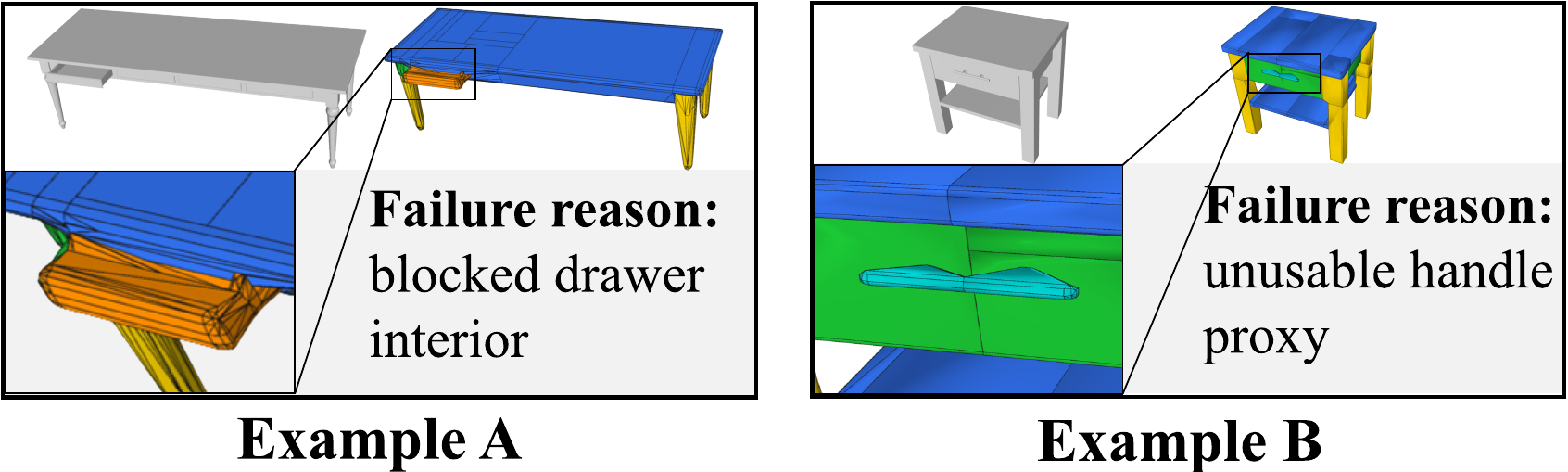}
	\caption{Examples of task-specific failures in collision meshes.}
	\label{fig:introExamples}
\end{figure}

\textbf{Challenge 1 (C1): Heterogeneity of collision meshes.}
Even within the same asset category, collision meshes vary in modeling style, decomposition granularity, and hull layout.
For instance, the \semlabel{handle} of one cup may decompose into two hulls, while that of another cup into five.
Whole-mesh batch-editing in existing tools are too coarse to express such part-level edits, while scripts over vertices, faces, hull indices, or fixed coordinates are too brittle because these low-level identifiers do not map consistently to semantic parts across heterogeneous meshes.
Thus, extractor synthesis must generalize beyond low-level geometric identifiers.

\textbf{Challenge 2 (C2): Insufficiency of semantic labels alone.}
Semantic labels provide a useful abstraction above raw geometry, and prior neuro-symbolic synthesis often uses neural perception labels as program predicates~\cite{yi2018neural,barnaby2023imageeye}.
However, labels alone are too coarse for collision mesh editing.
Collision mesh edits often target a subset of hulls sharing the same semantic label, such as handles attached to drawers rather than all handles, or only the bottommost hulls among all base hulls.
Thus, extractor logic must refine semantic labels using additional structural and spatial information.

\textbf{Challenge 3 (C3): Synthesis under perception noise.}
Despite advances in multimodal models, even state-of-the-art models still produce occasional errors (e.g., mistakenly annotating a \semlabel{body} hull as \semlabel{handle}).
This constitutes the primary bottleneck in program synthesis: since MeshForge must find an extractor consistent with \emph{all} demonstrations, even a single mislabeled hull can cause an otherwise correct extractor to be falsely rejected.
Without label correction, synthesis may be pushed to search for more complex extractors to accommodate the apparent contradiction, incurring overhead that far exceeds normal synthesis time.

These challenges motivate two design choices.
For \textbf{C1} and \textbf{C2}, we introduce a \textit{symbolic collision mesh} intermediate representation (IR) and a neuro-symbolic domain-specific language (DSL) for target selection.
The IR represents each hull by its semantic label, bounding box, and contact-graph relationships, enabling synthesis to reason over symbolic hulls rather than raw geometry.
The DSL then combines three extractor layers: \textit{semantic extractors} select labeled hulls, \textit{topological extractors} refine selections by attachment relationships, and \textit{geometric extractors} refine selections by spatial position.
For \textbf{C3}, inspired by the principle of abductive reasoning~\cite{dai2019bridging}, we design an \textit{Abductive Inference (ABI)} algorithm to handle sparse perception noise.
Unlike prior work that relies on users to provide corrections or disambiguation~\cite{barnaby2023imageeye,barnaby2025active}, ABI treats a near-miss extractor as evidence, infers the minimum label corrections that make the extractor consistent with \emph{all} demonstrations, and only asks the user to confirm them.

We evaluate MeshForge on a benchmark of 24 batch-editing tasks across eight 3D asset categories and 600 collision meshes.
MeshForge successfully synthesizes programs for 23 of 24 tasks, requiring 2.2 demonstrations and 3.5\,s of synthesis time on average. We summarize our contributions as follows:

\begin{itemize}[leftmargin=*]
	\item To the best of our knowledge, we are the first to identify collision mesh batch-editing, a new class of software-artifact maintenance problem, and to formalize it as a PBE task.

	\item We design a neuro-symbolic DSL built on our \textit{symbolic collision mesh}.
	      The DSL's three-layer extractor architecture enables reliable generalization across heterogeneous meshes.

	\item We present the ABI algorithm that mitigates the primary synthesis bottleneck caused by neural perception noise.

	\item We construct the first benchmark for collision mesh batch-editing, comprising 24 tasks across eight asset categories and 600 collision meshes, and evaluate MeshForge on it.
\end{itemize}

\section{Motivating Example}
\label{sec:motivating-example}
In this section, we use a table asset as a running example to illustrate the problem MeshForge solves.

Consider a developer working with a family of table assets in 3D software for robotic simulation.
The developer first generates collision meshes using a geometry-driven decomposition algorithm~\cite{wei2022approximate}.
However, these meshes do not fully support the intended interactions and thus require task-specific editing.
Since the asset family contains many table variants with different geometries and non-uniform orientations, editing them manually is tedious and error-prone.
The developer needs to:
\begin{enumerate*}[label=(\alph*)]
	\item remove task-irrelevant residual hulls;
	\item turn the fragmented tabletop into a stable support proxy for object placement;
	\item simplify each leg individually while preserving the legs as separate supports;
	\item clear the drawer interior so that objects can be placed inside; and
	\item make drawer-attached handles physically usable for robot interaction.
\end{enumerate*}

\begin{figure}[t]
	\centering
	\includegraphics[width=0.9\linewidth]{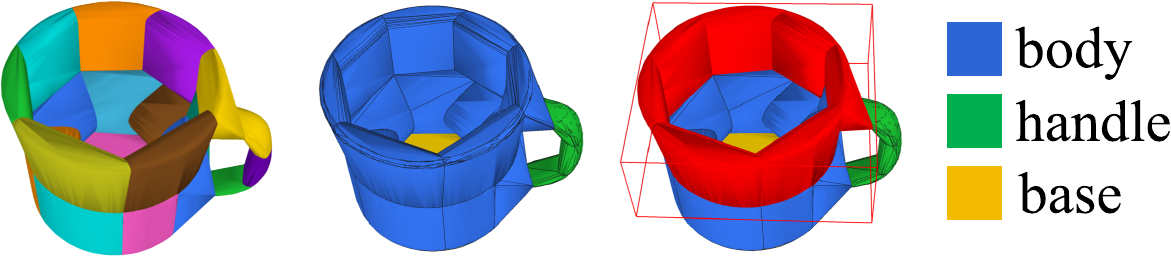}
	\caption{Representations of a collision mesh.
		Left: raw collision mesh.
		Middle: semantically annotated collision mesh.
		Right: selected hulls (red) and their aggregate bounding box.}
	\label{fig:collisionRepresentations}
\end{figure}

The developer could use MeshForge to automate these edits.
MeshForge first assigns semantic labels from a task-specific vocabulary (e.g., \semlabel{panel}, \semlabel{leg}, \semlabel{handle}, \semlabel{drawer}, and \semlabel{misc}), which the developer can inspect and correct in the GUI.
As illustrated in Fig.~\ref{fig:collisionRepresentations}, these labels lift the raw collision mesh into a representation where target hulls can be selected symbolically.
The developer then demonstrates the edits on a few representative meshes by selecting target hulls and choosing the corresponding editing actions.
Each demonstration provides examples of both \emph{what to edit} and \emph{how to edit it}: the action is recorded from the GUI, while the target-selection logic must generalize to table assets whose relevant hulls differ in number, shape, attachment, and orientation.
The developer may also specify orientation landmarks (one or two positive coordinate axes) so that directional operations are interpreted consistently.
From the demonstrations, MeshForge synthesizes reusable \texttt{\small extractor-action} rules and applies them to the remaining assets, which vary in size, appearance, and decomposition granularity.
The developer may then inspect the results, correcting isolated errors manually or providing additional demonstrations to re-synthesize if systematic errors appear.
\S\ref{sec:representation-and-dsl} defines the representation and DSL needed to express these rules; \S\ref{sec:approach} then describes how MeshForge synthesizes them from demonstrations.

\section{Symbolic Representation and DSL}
\label{sec:representation-and-dsl}
The PBE problem in MeshForge focuses on synthesizing target-selection logic that generalizes across collision meshes; the corresponding actions are recorded directly from demonstrations.
This requires both a representation that exposes semantic, spatial, and attachment information (\S\ref{sec:symbolic-collision-mesh}), and a DSL for reusable extractor-action rules (\S\ref{sec:dsl}).
We then instantiate the DSL on the motivating example (\S\ref{sec:running-example}).

\subsection{Symbolic Collision Mesh}
\label{sec:symbolic-collision-mesh}
A raw collision mesh \( M \) is a finite set of convex hulls, where each hull \( h \in M \) is defined by its vertices and faces.
Raw hulls are directly editable, but they do not expose the signals needed for reusable target selection: semantic roles, spatial positions, and attachment relationships.
To bridge this gap, we introduce the \textit{symbolic collision mesh}, an IR that augments raw hulls with this information while preserving a one-to-one link to the editable hulls.
We formalize it as follows.

\begin{definition}[Symbolic collision mesh]
	Given a raw collision mesh $M$, its \textit{symbolic collision mesh} is an undirected graph $\hat{M} = (S, E)$, where each node $s_g \in S$ is a \textit{symbolic hull} that corresponds uniquely to a raw hull $h_g \in M$ (indexed by $g$).
	Each symbolic hull $s_g$ carries two attributes: (1) a semantic label \( s_g.l \in \Sigma \), where $\Sigma$ is a finite task-specific label vocabulary; and (2) an axis-aligned bounding box (AABB) $s_g.b = (c_g, \mathbf{r}_g)$, where center $c_g \in \mathbb{R}^3$ and half-extents $\mathbf{r}_g \in \mathbb{R}^3_{>0}$ along each axis ($r_{g,\alpha}$ denotes the half-extent along axis~$\alpha \in \{X,Y,Z\}$).
	The edge set $E \subseteq \binom{S}{2}$ encodes hull adjacency:
	$\{s_u, s_v\} \in E$ if and only if
	$
		\mathrm{dist}(h_u, h_v) \;\leq\; \epsilon \cdot \mathrm{diag}(B(M)),
	$
	where $B(M) = \mathrm{AABB}(M)$ is the tightest AABB enclosing $M$, $\mathrm{diag}(B(M))$ is its spatial diagonal length, $\mathrm{dist}(h_u, h_v) = \min_{\mathbf{p} \in h_u,\, \mathbf{q} \in h_v} \|\mathbf{p} - \mathbf{q}\|_2$ is the minimum Euclidean distance between $h_u$ and $h_v$, $\mathbf{p}$ and $\mathbf{q}$ are points on them, and $\epsilon > 0$ is a dimensionless threshold (see \S\ref{sec:implementation}).
	We refer to the adjacency structure defined by $E$ as the contact graph.
	We write $\mathcal{N}(s_g) = \{s_{g'} \mid \{s_g, s_{g'}\} \in E\}$ for the neighbor set of $s_g$.
\end{definition}

The label \( s_g.l \) is annotated by the pre-trained model.
Users can customize $\Sigma$ so that the label granularity fits their editing needs.
Fig.~\ref{fig:collisionRepresentations} illustrates how MeshForge lifts a raw collision mesh into a symbolic representation: hulls are assigned task-specific labels.
The label supports semantic selection, the AABB supports geometric selection, and the contact graph supports attachment-based selection.
These three forms of information provide the basis for the DSL introduced next.

\begin{figure}[t]
	\centering
	\small
	\[
		\begin{array}{r@{\;}c@{\;}l}
			\mathcal{P}               & ::= & \Phi;\;\, r_1;\, r_2;\, \ldots;\, r_n                                          \\[3pt]

			\Phi                      & ::= & \op{Identity}
			\mid \op{Orient}(\pi)
			\mid \op{Orient}(\pi_1,\,\pi_2)                                                                                  \\[3pt]

			\pi                       & ::= & \langle \mathcal{E}_{\text{base}},\, \mathcal{E}_{\text{base}} \rangle         \\[3pt]

			r                         & ::= & \mathcal{E} \mapsto \mathcal{A}
			\mid \mathcal{E} \mapsto \op{Each}(\mathcal{A})                                                                  \\[3pt]

			\mathcal{E}_{\text{base}} & ::= & \op{All} \mid \op{Is}(a)                                                       \\[3pt]

			\mathcal{E}               & ::= & \mathcal{E}_{\text{base}}
			\mid \op{Attach}(\mathcal{E}_{\text{base}},\, \mathcal{E}_{\text{base}})
			\mid \op{Geo}(\mathcal{E}_{\text{base}},\, \mathit{d})                                                           \\[2pt]
			                          &     & \mid \op{Union}(\mathcal{E},\mathcal{E})
			\mid \op{Intersect}(\mathcal{E},\mathcal{E})
			\mid \op{Diff}(\mathcal{E},\mathcal{E})                                                                          \\[3pt]

			\mathcal{A}               & ::= & \op{Delete} \mid \op{Replace}(g) \mid \op{Add}(\mathit{d},\, \tau)             \\[2pt]
			                          &     & \mid \op{Merge} \mid \op{Scale}(\mathit{axis},\,\lambda)                       \\[3pt]

			\mathit{axis}             & ::= & \textsf{all} \mid X \mid Y \mid Z                                              \\[3pt]

			\mathit{d}                & ::= & {+}X \mid {-}X \mid {+}Y \mid {-}Y \mid {+}Z \mid {-}Z                         \\[3pt]
			g                         & ::= & \tau \mid \rho                                                                 \\[3pt]

			\rho                      & ::= & \textsf{box} \mid \textsf{cylinder} \mid \textsf{capsule} \mid \textsf{sphere} \\[3pt]

			a                         & \in & \Sigma \quad \lambda \in \mathbb{R}_{>0} \quad \tau:\ \text{hull template}
		\end{array}
	\]
	\caption{MeshForge DSL for collision mesh editing.}
	\label{fig:dsl}
\end{figure}

\subsection{DSL for Batch-Editing Programs}
\label{sec:dsl}
We present the DSL in three parts: extractors (target selection), actions (editing), and program structure.

\subsubsection{Extractors: What to Edit}
\label{sec:dsl-extractors}
Extractors select the target hull subset for each rule.
Since edits may depend on a hull's spatial position or attachment relationships, semantic labels alone are too coarse.
MeshForge therefore provides three complementary extractor layers.

Formally, each extractor denotes a hull set $\llbracket \cdot \rrbracket_{\hat{M}} \subseteq S$, with denotational semantics given in Fig.~\ref{fig:semantics}.
\textit{Semantic extractors} select hulls by semantic label $a \in \Sigma$, or select all hulls using \op{All}.
\textit{Topological extractors} refine base selections using contact-graph attachment:
\op{Attach}($\mathcal{E}_{\text{base},1}, \mathcal{E}_{\text{base},2}$) selects hulls from the first base selection that are adjacent in the contact graph to at least one hull in the second.
\textit{Geometric extractors} refine a base selection using spatial extremity:
\op{Geo}($\mathcal{E}_{\text{base}}, d$) selects hulls whose extreme coordinate $\mathrm{E}_d(s_g)$ along direction $d$ is within $\delta\cdot\mathrm{D}(Q)$ of the extreme boundary of the base selection $Q$, where $\mathrm{E}_d(s_g)$ is the extreme coordinate of $s_g$'s AABB in direction $d$, $\mathrm{D}(Q)$ is the diagonal of the aggregate AABB of $Q$, and $\delta > 0$ is a dimensionless threshold (see \S\ref{sec:implementation}).
\op{Union}, \op{Intersect}, and \op{Diff} express standard set semantics over the hull sets denoted by their operands.
We intentionally restrict \op{Attach} and \op{Geo} to base selections to avoid overfitting to specific decomposition granularities.
For example, if nested geometric refinements were allowed, expressions such as \op{Geo}(\op{Geo}(\op{Is}(\semlabel{body}), $+Y$), $-X$) could overfit to hulls produced by a particular decomposition pattern.
Set operators (\op{Union}, \op{Intersect}, \op{Diff}) compose these layers into reusable extraction logic across heterogeneous meshes.

\begin{figure}[t]
	\centering
	\small
	\[
		\begin{array}{r@{\;}c@{\;}l}
			\multicolumn{3}{c}{\textbf{Orientation semantics}}                                                    \\[2pt]
			\hline                                                                                                \\[-6pt]

			\llbracket \op{Identity} \rrbracket(\hat{M})
			 & = &
			\hat{M}                                                                                               \\[3pt]

			\llbracket \op{Orient}(\pi) \rrbracket(\hat{M})
			 & = &
			\textsc{Rot}(\hat{M},\, \textsc{Axis}(\pi))                                                           \\[3pt]

			\llbracket \op{Orient}(\pi_1,\, \pi_2) \rrbracket(\hat{M})
			 & = &
			\textsc{Rot}(\hat{M},\, \textsc{Axis}(\pi_1),\, \textsc{Axis}(\pi_2))                                 \\[5pt]

			\multicolumn{3}{c}{\textbf{Extractor semantics}}                                                      \\[2pt]
			\hline                                                                                                \\[-6pt]

			\llbracket \op{All} \rrbracket_{\hat{M}}
			 & = &
			S                                                                                                     \\[3pt]

			\llbracket \op{Is}(a) \rrbracket_{\hat{M}}
			 & = &
			\{s_g \in S \mid s_g.l = a\}                                                                          \\[3pt]

			\llbracket \op{Attach}(\mathcal{E}_{\text{base},1},\, \mathcal{E}_{\text{base},2}) \rrbracket_{\hat{M}}
			 & = &
			\{s_g \in Q_1 \mid                                                                                    \\
			 &   &
			\hspace{0.5em} \exists s_{g'} \in Q_2:\{s_g,s_{g'}\}\in E\}                                           \\[3pt]

			\llbracket \op{Geo}(\mathcal{E}_{\text{base}},\, d) \rrbracket_{\hat{M}}
			 & = &
			\{s_g \in Q \mid \max_{s_{g'} \in Q}\,                                                                \\
			 &   &
			\hspace{0.5em}  \mathrm{E}_d(s_{g'}) - \mathrm{E}_d(s_g) \leq \delta\cdot\mathrm{D}(Q)\}              \\[3pt]

			\llbracket \op{Union}(\mathcal{E}_1, \mathcal{E}_2) \rrbracket_{\hat{M}}
			 & = &
			\llbracket \mathcal{E}_1 \rrbracket_{\hat{M}} \cup \llbracket \mathcal{E}_2 \rrbracket_{\hat{M}}      \\[3pt]

			\llbracket \op{Intersect}(\mathcal{E}_1, \mathcal{E}_2) \rrbracket_{\hat{M}}
			 & = &
			\llbracket \mathcal{E}_1 \rrbracket_{\hat{M}} \cap \llbracket \mathcal{E}_2 \rrbracket_{\hat{M}}      \\[3pt]

			\llbracket \op{Diff}(\mathcal{E}_1, \mathcal{E}_2) \rrbracket_{\hat{M}}
			 & = &
			\llbracket \mathcal{E}_1 \rrbracket_{\hat{M}} \setminus \llbracket \mathcal{E}_2 \rrbracket_{\hat{M}} \\[5pt]

			\multicolumn{3}{c}{\textbf{Action semantics}}                                                         \\[2pt]
			\hline                                                                                                \\[-6pt]

			\llbracket \mathcal{E} \mapsto \op{Delete} \rrbracket(\hat{M})
			 & = &
			\hat{M} \setminus C                                                                                   \\[3pt]

			\llbracket \mathcal{E} \mapsto \op{Replace}(g) \rrbracket(\hat{M})
			 & = &
			(\hat{M} \setminus C) \oplus \gamma(g,\; B(C))                                                        \\[3pt]

			\llbracket \mathcal{E} \mapsto \op{Add}(d,\, \tau) \rrbracket(\hat{M})
			 & = &
			\hat{M} \oplus \textsc{Place}(\tau,\; B(C),\; d)                                                      \\[3pt]

			\llbracket \mathcal{E} \mapsto \op{Merge} \rrbracket(\hat{M})
			 & = &
			(\hat{M} \setminus C) \oplus \mathrm{CH}\big(\textstyle\bigcup_{s_g \in C} h_g\big)                   \\[3pt]

			\llbracket \mathcal{E} \mapsto \op{Scale}(\mathit{axis},\,\lambda) \rrbracket(\hat{M})
			 & = &
			(\hat{M} \setminus C) \oplus \sigma_{\mathit{axis},\lambda}(C)
		\end{array}
	\]
	\caption{Semantics of the MeshForge DSL.}
	\label{fig:semantics}
\end{figure}

\subsubsection{Actions: How to Edit}
\label{sec:dsl-actions}
Editing actions transform the selected hulls.
All actions are defined at the hull level, since physics engines require convex collision meshes and arbitrary sub-hull edits (e.g., vertex-level edits) may violate convexity.
MeshForge supports five hull-level actions: \op{Delete}, \op{Replace}, \op{Add}, \op{Merge}, and \op{Scale}.
For \op{Replace}, the inserted geometry $g$ can be either a user-provided hull template $\tau$ or a built-in convex primitive $\rho$ (box, cylinder, capsule, or sphere).
Users can model templates in dedicated 3D modeling software~\cite{blender2023} and import them into MeshForge.

The lower part of Fig.~\ref{fig:semantics} formalizes these action semantics.
Given a rule $\mathcal{E} \mapsto \mathcal{A}$ applied to $\hat{M}$, MeshForge evaluates the extractor to obtain the target hull set $C = \llbracket \mathcal{E} \rrbracket_{\hat{M}}$ and its AABB $B(C)$.
\op{Delete} removes $C$.
\op{Replace}($g$) removes $C$ and inserts $\gamma(g, B(C))$, where $\gamma$ aligns $g$ so that its AABB coincides with $B(C)$.
\op{Add}($d,\tau$) retains $C$ and inserts $\textsc{Place}(\tau, B(C), d)$, which scales and positions $\tau$ flush against $B(C)$ along direction $d$; unlike \op{Replace}, it accepts only a template $\tau$, since a primitive placed along $d$ would have an underdetermined orientation.
\op{Merge} replaces $C$ with the convex hull enclosing the geometry of all hulls in $C$.
\op{Scale}($\mathit{axis},\,\lambda$) scales $C$ as a group by factor $\lambda$ either along all axes or along the specified axis $\mathit{axis}$.
All actions preserve hull convexity by construction, and contact-graph edges are recomputed after each action to stay consistent with the edited mesh.

\subsubsection{Program Structure}
\label{sec:dsl-structure}
The DSL grammar is shown in Fig.~\ref{fig:dsl}.
A batch-editing program composes extractors and actions into ordered rules.
Each rule has the form $\mathcal{E} \mapsto \mathcal{A}$, or $\mathcal{E} \mapsto \op{Each}(\mathcal{A})$, which partitions the selected set $C$ into connected components in the contact graph and applies $\mathcal{A}$ independently to each.
This component-wise form is useful when one extractor selects multiple disconnected but semantically equivalent parts, such as four chair legs.
A program $\mathcal{P}$ begins with an orientation directive $\Phi$, followed by rules that are executed sequentially in the order they appear in $\mathcal{P}$.

Since geometric extractors and directional actions depend on coordinate directions, we introduce an orientation normalization mechanism before applying the edits.
Specifically, \op{Orient} uses one or two user-provided landmark pairs to define canonical axes before rule execution, allowing directional operators (e.g., \op{Geo}($\cdot,d$), \op{Add}($d,\tau$)) to be interpreted consistently across assets.
\op{Identity} leaves the mesh in its original coordinate frame when no orientation normalization is needed.
A single axis is sufficient when only one directional reference is needed (e.g., aligning the vertical axis of a bottle), whereas two axes determine a full canonical frame; the third axis is implied by the right-hand rule.
As shown in Fig.~\ref{fig:semantics}, \op{Orient} rotates the underlying hull geometry and recomputes AABBs.

\subsection{Running Example}
\label{sec:running-example}
\begin{figure}[t]
	\centering
	\includegraphics[width=\linewidth]{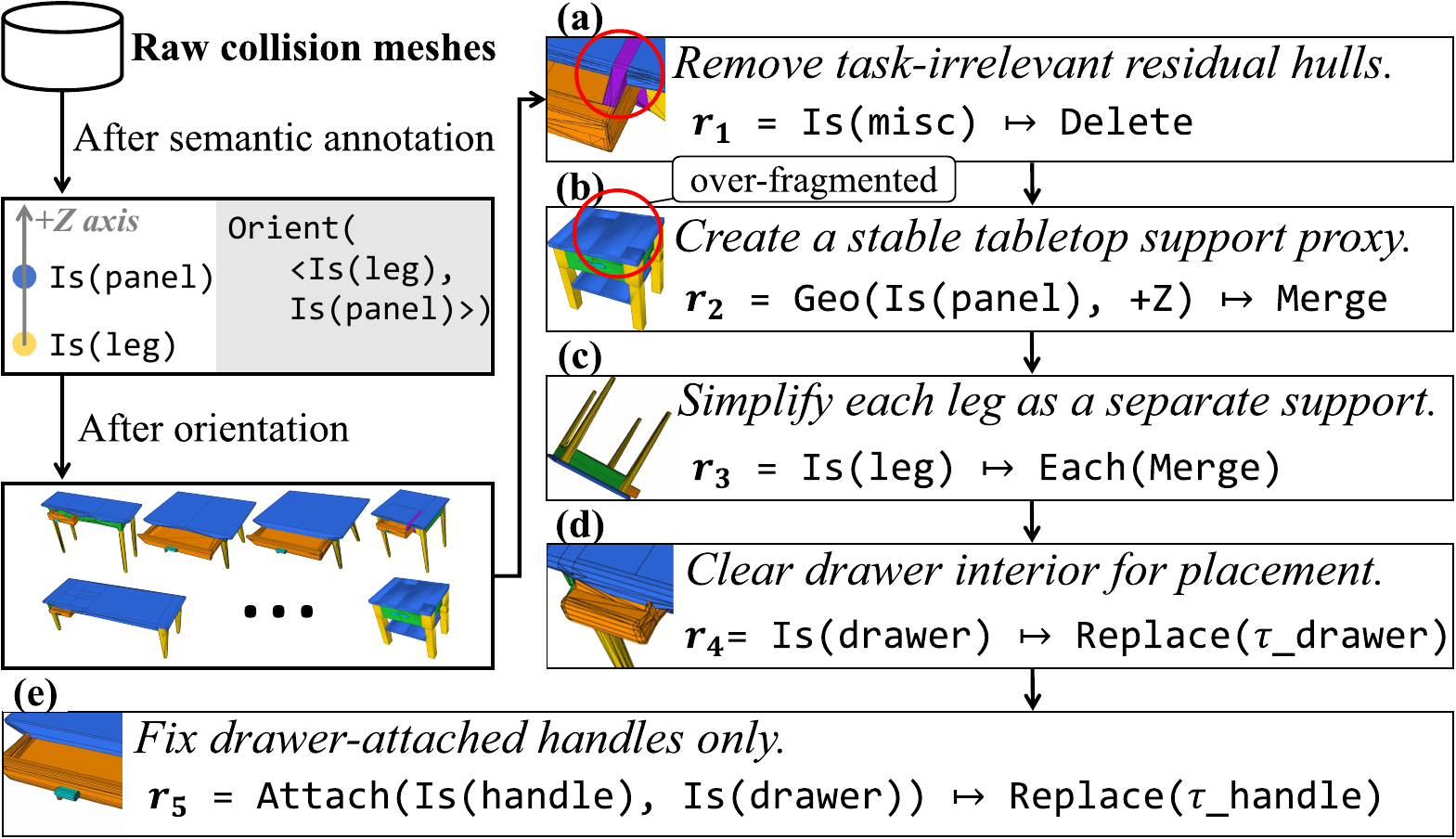}
	\caption{Running example of the DSL on table collision meshes.}
	\label{fig:runningExample}
\end{figure}

Fig.~\ref{fig:runningExample} shows the synthesized DSL program for the motivating example introduced in \S\ref{sec:motivating-example}.
The orientation directive first normalizes the input meshes by defining the direction from the center of \semlabel{leg} hulls to the center of \semlabel{panel} hulls as the positive vertical axis ($+Z$).
The synthesized rules then express the five edits.
This example illustrates how the DSL combines orientation normalization, semantic/topological/geometric extractors, component-wise application, and hull-level actions to express a reusable batch-editing program.

\section{Our Approach: MeshForge}
\label{sec:approach}

\subsection{Overview of the Pipeline}
\begin{figure}[t]
	\centering
	\includegraphics[width=\linewidth]{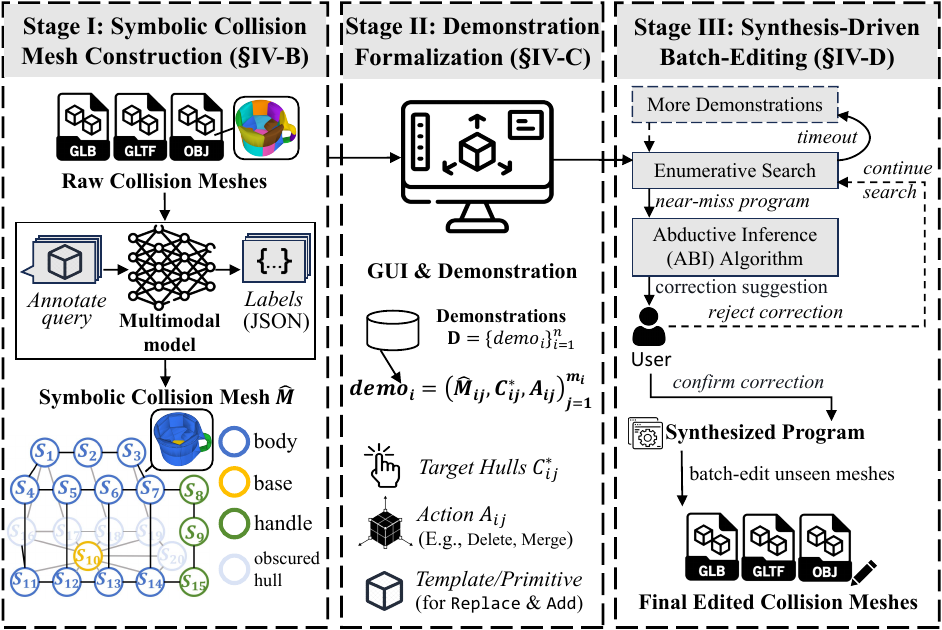}
	\caption{Overview of the MeshForge pipeline.}
	\label{fig:pipeline}
\end{figure}

Fig.~\ref{fig:pipeline} illustrates the pipeline of MeshForge.
Given raw collision meshes that share the same editing intent, MeshForge constructs symbolic collision meshes, records user demonstrations as formal editing examples, synthesizes reusable rules, and applies the synthesized program to non-demonstration meshes.
Users demonstrate \emph{what} hulls to edit and \emph{how} to edit them, while MeshForge synthesizes the reusable extractor logic across the batch.
To design this pipeline, we follow four SE principles: (1) \textbf{Modularity}: each component is independently replaceable, easing future extension; (2) \textbf{Inspectability}: both the symbolic IR and synthesized programs are human-readable, letting developers audit intermediate results; (3) \textbf{Legality}: all editing operations preserve collision mesh legality, ensuring physics-engine compatibility; and (4) \textbf{Controllability}: users can intervene at multiple points (e.g., correcting annotations, supplying additional demonstrations, and refining edited meshes).

\subsection{Stage I: Symbolic Collision Mesh Construction}
\label{sec:stage1}
Users provide raw collision meshes sharing the same editing intent, and MeshForge constructs a symbolic collision mesh IR (see \S\ref{sec:symbolic-collision-mesh}) for each.
Construction first assigns semantic labels with a neural model, then augments each hull with the geometric and topological information that completes the IR (AABBs and contact-graph edges).
For label assignment, we render a set of orthographic views that highlight each hull in a distinct color.
Given the user-specified label vocabulary $\Sigma$, the neural model is prompted with these views to label each hull based on its visual appearance and apparent spatial context.
MeshForge provides a GUI for users to inspect and manually correct annotations if necessary.
The resulting symbolic collision mesh serves as the input to Stage~II.

\subsection{Stage II: Demonstration Formalization}
\label{sec:stage2}
Stage II converts GUI edits into the formal examples used by synthesis.
Instead of expressing requirements in natural language or writing DSL programs, users demonstrate edits through a GUI.
These interactions resemble standard 3D modeling tools that developers already use, reducing cognitive burden.
In practice, after Stage~I, users select a small number of representative meshes for demonstration (typically 2--3, depending on task complexity).
Although more demonstration meshes improve generalization, users need not provide more upfront.
Instead, if the synthesized program proves insufficient in Stage~III, users can incrementally add more demonstrations and re-synthesize, a process we refer to as iterative refinement.
To address orientation misalignment in the input dataset, users may optionally specify orientation landmarks through the same GUI, as illustrated in \S\ref{sec:running-example}.
Each demonstration consists of one or more editing steps on each selected mesh.
When multiple steps are demonstrated, MeshForge records them as an ordered sequence of rules.
For each rule, the demonstration provides the target hulls and editing action, while leaving the reusable extractor to be synthesized in Stage~III.
We represent the demonstration of the $i$-th rule as $\mathit{demo}_i = \{(\hat{M}_{ij},\, C_{ij}^*,\, A_{ij})\}_{j=1}^{m_i}$, where $m_i$ is the number of demonstration meshes for rule $i$, $\hat{M}_{ij}$ is the intermediate symbolic collision mesh of the $j$-th demonstration mesh after applying $\Phi$ and rules $r_1, \ldots, r_{i-1}$ ($\hat{M}_{1j}$ is the initial mesh after applying $\Phi$), $C_{ij}^* \subseteq S_{ij}$ is the user-selected target hull set in $\hat{M}_{ij}$, and $A_{ij}$ is the demonstrated editing action.
For \op{Replace}, MeshForge records the inserted geometry $g$; for \op{Add}, it records the template $\tau$ and placement direction $d$ as part of the rule.

\subsection{Stage III: Synthesis-Driven Batch-Editing}
\label{sec:stage3}
Given demonstrations $\mathbf{D}$ from Stage~II, MeshForge searches for extractor logic that makes each recorded action reusable.
The search first seeks a DSL program consistent with $\mathbf{D}$.
When a complete candidate fails to match, MeshForge invokes ABI to test whether sparse label corrections explain the mismatch, returning a repaired extractor or resuming the search.
Once a program matches the demonstrations, it is applied to every remaining mesh.
Users then inspect the editing results.
If satisfactory, the process concludes.
Otherwise, users may manually correct a few failed cases or provide additional demonstrations to trigger iterative refinement (\S\ref{sec:stage2}).
Below, we formalize the synthesis problem, describe the search algorithm (\S\ref{sec:synthesis-algorithm}), and explain how MeshForge abductively corrects neural labels (\S\ref{sec:neural-noise-handling}).

\begin{definition}[Program Synthesis]
	Given demonstrations $\mathbf{D} = \{\mathit{demo}_i\}_{i=1}^n$ where $\mathit{demo}_i = \{(\hat{M}_{ij},\, C_{ij}^*,\, A_{ij})\}_{j=1}^{m_i}$, synthesize a program $\mathcal{P} = \Phi;\, r_1;\, \ldots;\, r_n$ with each rule $r_i = \mathcal{E}^{(i)} \mapsto \mathcal{A}_i$.
	Each $\mathcal{E}^{(i)}$ satisfies:
	\[
		\forall j \in \{1,\ldots,m_i\}:\quad
		\llbracket \mathcal{E}^{(i)} \rrbracket_{\hat{M}_{ij}} = C_{ij}^*
	\]
	Since the orientation directive $\Phi$ and actions $\mathcal{A}_i$ are recorded from the demonstrations, synthesis only searches for $\mathcal{E}^{(i)}$.
	During synthesis, MeshForge processes rules sequentially: for each $i$, it reconstructs the intermediate meshes by applying $\Phi$ and the previously synthesized rules $r_1,\ldots,r_{i-1}$, and then searches for $\mathcal{E}^{(i)}$ consistent with $C_{ij}^*$ across all demonstrations.
\end{definition}

\subsubsection{Batch-Editing Program Synthesis}
\label{sec:synthesis-algorithm}
To infer the extractor for each rule $r_i$, MeshForge performs top-down enumerative search over the DSL, conditioned on the intermediate meshes from previously synthesized rules (Algorithm~\ref{alg:extractor}).
A \textit{partial program} is an extractor expression with one or more holes ($\square$), where each $\square$ marks a sub-expression still to be instantiated. \textsc{Complete}($p$) holds when no holes remain.
The worklist $W$ is initialized with a single hole (Line~1), and \textsc{PopMin} dequeues the candidate of lowest syntactic complexity, measured as AST node count (Line~3).
For each complete candidate, \textsc{Match} checks consistency with all demonstrations (Line~5).
If it passes, $p$ is returned immediately (Line~6).
If \textsc{Match} fails, \textsc{AbductiveInference} is invoked to determine whether the mismatch stems from a flawed program or from sparse neural perception noise (Lines~8--10).
For incomplete candidates, \textsc{EnumerateCandidates} fills the next hole and returns only feasible refinements (Line~13).
Synthesis terminates when a valid or repaired extractor is found, or upon timeout.

\begin{algorithm}[t]
	\caption{\textsc{SynthesizeExtractor}}
	\label{alg:extractor}
	\small
	\begin{algorithmic}[1]
		\REQUIRE \(\mathit{demo}_i\); \(L\) (neural semantic annotations)
		\ENSURE Extractor \(\mathcal{E}^{(i)}\), or \((\mathcal{E}^{(i)},\, R)\) with label corrections, or \(\bot\)

		\STATE \(W \gets \{\Box\}\) \hfill\textit{/* worklist of candidate extractor programs */}
		\WHILE{\(W \neq \emptyset\)}
		\STATE \(p \gets \textsc{PopMin}(W)\) \hfill\textit{/* dequeue simplest candidate */}
		\IF{\(\textsc{Complete}(p)\)}
		\IF{\(\textsc{Match}(p,\, \mathit{demo}_i)\)}
		\STATE \textbf{return} \(p\) \hfill\textit{/* the synthesized $\mathcal{E}^{(i)}$ */}
		\ENDIF
		\STATE \(R \gets \textsc{AbductiveInference}(p,\, \mathit{demo}_i,\, L)\)
		\IF{\(R \neq \bot\) \textbf{and} \(\textsc{UserConfirms}(R)\)}
		\STATE \textbf{return} \((p,\, R)\)
		\ENDIF
		\ELSE
		\STATE \(W \gets W \cup \textsc{EnumerateCandidates}(p,\, \mathit{demo}_i)\)
		\ENDIF
		\ENDWHILE
		\STATE \textbf{return} \(\bot\)
	\end{algorithmic}
\end{algorithm}

\subsubsection{Abductive Correction of Neural Labels}
\label{sec:neural-noise-handling}
Despite recent advances in multimodal models, neural perception noise remains inevitable.
Although perception noise is sparse (see \S\ref{sec:implementation}), its impact on demonstration meshes is severe: even one mislabeled hull can reject an otherwise correct program (a \textit{near-miss program}) and push synthesis toward increasingly complex alternatives, causing synthesis to fail or degrade.

The fundamental difficulty is the absence of a repair oracle: no system can independently determine which annotations are wrong.
Prior work addresses perception noise by requiring user correction or disambiguation queries~\cite{barnaby2023imageeye,barnaby2025active}, ultimately depending on additional user input.
We take a different approach: inspired by abductive reasoning~\cite{dai2019bridging}, we treat a near-miss program $p$ as evidence, where the mismatch between $p$'s output and the user-selected hulls identifies suspect semantic annotations.
We then infer the minimum label corrections that make $p$ consistent with all demonstrations.
The user only needs to \emph{confirm} the inferred corrections rather than \emph{provide} them.

The ABI algorithm proceeds in two phases (Algorithm~\ref{alg:abductive-inference}).
First, it constructs the suspect set $U$ from the symmetric difference between the current candidate extractor $p$'s output and the user-selected target hull set (Lines~1--4), where $\llbracket p\rrbracket_{\hat{M}_j,L}$ denotes evaluating $p$ on $\hat{M}_j$ under label assignment $L$.
These hulls are suspects whose annotations may explain why $p$ disagrees with the demonstrations.
Then, in the repair phase, the algorithm enumerates candidate relabeling functions via $\textsc{CandidateRepairs}(U, p, L, \eta)$ (Line~6), which returns assignments that change at most $\eta$ hull labels per demo using semantic labels relevant to $p$.
For each candidate, \textsc{ApplyRepair} applies the relabeling to $\mathit{demo}_i$ and \textsc{Match} checks full consistency (Line~7).
The algorithm then computes repair cost and retains the minimum-cost repair as $R^*$ (Lines~8--12).

This algorithm remains tractable because annotation errors are sparse (see \S\ref{sec:implementation}), keeping $U$ small.
The search is further constrained by a noise threshold $\eta$ that upper-bounds the number of allowed corrections.
If many suspects require correction, it likely indicates a more systemic issue (e.g., a flawed synthesized program), and the algorithm returns $\bot$ rather than proposing an overwhelming number of corrections.
We further discuss the value of $\eta$ in \S\ref{sec:implementation}.
When $R^* \neq \bot$, MeshForge surfaces the corrections for user confirmation rather than applying them automatically, allowing synthesis to terminate early with a repaired extractor.

\begin{algorithm}[t]
	\caption{\textsc{AbductiveInference}}
	\label{alg:abductive-inference}
	\small
	\begin{algorithmic}[1]
		\REQUIRE $p$; $\mathit{demo}_i$; $L$ (semantic annotations); $\eta$ (noise threshold)
		\ENSURE Minimal repair set $R^*$, or $\bot$

		\STATE $U \gets \emptyset$ \hfill\textit{/* suspect hulls whose labels may need correction */}
		\FOR{\textbf{each} $(\hat{M}_j,\, C^*_j, \cdot) \in \mathit{demo}_i$}
		\STATE $U \gets U \cup \left(\llbracket p \rrbracket_{\hat{M}_j,\,L} \;\triangle\; C^*_j\right)$
		\ENDFOR
		\STATE $R^* \gets \bot$;\quad $c^* \gets \infty$
		\FOR{\textbf{each} $f \in \textsc{CandidateRepairs}(U,\, p,\, L,\, \eta)$}
		\IF{$\textsc{Match}(p,\; \textsc{ApplyRepair}(\mathit{demo}_i,\, f))$}
		\STATE $c \gets |\{h \in U : f(h) \neq L(h)\}|$
		\IF{$c < c^*$}
		\STATE $R^* \gets \{h \mapsto f(h) \mid h \in U,\; f(h) \neq L(h)\}$
		\STATE $c^* \gets c$
		\ENDIF
		\ENDIF
		\ENDFOR
		\STATE \textbf{return} $R^*$
	\end{algorithmic}
\end{algorithm}

\section{Implementation}
\label{sec:implementation}
We conduct all experiments on a desktop machine with a 2.6~GHz 13th-Gen Intel Core i5 CPU and 16~GB RAM.

\head{Neural Perception.}
Effective hull annotation imposes two requirements on the neural model: (1) high label assignment accuracy, and (2) support for open-world label vocabularies at flexible granularity.
We exclude supervised methods (e.g., Pointnext~\cite{qian2022pointnext}) because they do not satisfy requirement~(2).
Among the remaining candidates, we evaluate three representative paradigms: open-world 3D segmentation (Find3D~\cite{ma2025find}), open-source vision-language model (VLM) (Qwen3-VL~\cite{Qwen3-VL}), and proprietary VLM (GPT-4o-2024-11-20~\cite{openai2024gpt4o}).
We evaluate annotation accuracy on 200 randomly selected collision meshes disjoint from the evaluation benchmark, measuring per-hull accuracy against reference annotations under each task's label vocabulary; any invalid or out-of-vocabulary predictions are counted as incorrect.
The reference annotations achieve substantial to almost perfect inter-annotator agreement across tasks (Fleiss'~$\kappa > 0.85$).
GPT-4o achieves the highest accuracy (0.92), compared with Find3D (0.61) and Qwen3-VL (0.76), and is thus adopted, at an average cost of \$0.0224 per mesh.
All VLMs are configured with temperature 0. To check response consistency, we repeat each GPT-4o annotation five times on the sampled meshes and find no significant variation (Friedman test, $p = 0.54$).
This indicates that GPT-4o annotations are stable under repeated queries in our setting, so we use a single annotation query per mesh.

\head{Hyper-parameters Setting.}
We tune $\epsilon$, $\delta$, and $\eta$ empirically on the same 200-mesh sample above,  which we partition into a calibration set (150 meshes) and a test set (50 meshes).
For $\epsilon$, we manually annotate whether nearby hull pairs should be considered adjacent and set \(\epsilon = 0.002\), which achieves 100\% accuracy on both sets.
This indicates that a fixed threshold suffices.
For $\delta$ (\op{Geo}), we manually annotate boundary hulls along each axis.
We set $\delta = 0.05$, achieving 96\% accuracy on the test set.
For $\eta$ (ABI), we reuse the 200-mesh sample, which has both reference and VLM-predicted labels.
For each mesh, we count the number of mislabeled hulls.
The per-mesh mislabel count has a 95th percentile of 2, indicating that annotation errors are sparse.
We therefore set $\eta = 2$, which also limits user confirmation to at most two label corrections, keeping user overhead minimal.

\section{Evaluation}
\label{sec:evaluation}
We evaluate MeshForge via three research questions (RQs):
\begin{itemize}[leftmargin=*]
	\item \textbf{RQ1 (Effectiveness):} Can MeshForge synthesize batch-editing programs across heterogeneous collision meshes?
	\item \textbf{RQ2 (Comparison):} How does MeshForge compare against existing approaches, quantitatively and qualitatively?
	\item \textbf{RQ3 (Ablation Study):} How do individual components contribute to its overall effectiveness and efficiency?
\end{itemize}

\subsection{Experimental Setup}
\label{sec:experimental-setup}
\head{Benchmark.}
To construct a benchmark for MeshForge evaluation, we first collect discussions and documentation related to collision meshes in 3D software, including posts from online developer forums, official documentation, and papers.
We retain all sources describing concrete collision editing needs, filtering out only irrelevant ones (e.g., those that explain general collision concepts).
We then map the retained sources to a set of editing motifs.
Finally, we instantiate these motifs on assets from public datasets~\cite{yu2019partnet,calli2015ycb}.
Overall, the benchmark covers eight asset categories: cup, kettle, knife, chair, dispenser, door, lamp, and table.
For each 3D asset, to assess generality, we generate three collision mesh variants using CoACD~\cite{wei2022approximate}, a widely used convex decomposition algorithm, with coarse, medium, and fine decomposition settings.
We instantiate three batch-editing tasks of increasing difficulty (easy, medium, and hard) for each asset category. The difficulty is determined by the number of editing rules and the complexity of the extractor logic.
The resulting benchmark comprises 24 batch-editing tasks across eight asset categories, totaling 200 assets and 600 collision meshes.
For each task, we encode a reference DSL program for evaluating synthesis behavior.

\head{Protocol.}
Five developers with over three years of experience in 3D software development use MeshForge to complete all evaluation tasks independently.
None of them was involved in the design of MeshForge, and none had prior experience with the system.
Before evaluation, they are introduced to MeshForge's workflow and practice on examples not included in the benchmark.
For each task, a developer selects $k$ representative meshes as demonstrations through the GUI, and MeshForge synthesizes a DSL program from these demonstrations.
The developer may inspect the synthesized program's results on the non-demonstration meshes and add demonstrations when systematic failures are observed.
Each task starts with $k=1$, and increases $k$ until the developer accepts the synthesized program or $k$ reaches the maximum of 5.
The final accepted program is evaluated on the non-demonstration meshes, with each synthesis round subject to a 120-second timeout.
All tasks use predefined hull templates, so template construction is not evaluated.
We strictly follow the standard PBE evaluation protocol~\cite{rolim2017learning,meng2013lase,dong2022webrobot}, partitioning each task's meshes into a demonstration set and a non-demonstration set based on user selection.
Unlike supervised learning, PBE has no fixed train/test split: the user chooses which meshes to demonstrate on. Imposing an artificial split could place rare variants entirely in the test set, where a failure would reflect a missing demonstration, rather than a true generalization error.
The label vocabulary $\Sigma$ is freely specified by the user (cost subsumed in user interaction time). We found all five developers converged to semantically equivalent vocabularies per task (averaging 4.2 labels), suggesting the required semantic granularity is intuitive and unambiguous.

\head{Metrics.}
We report \textit{Synthesis Success Rate (SSR)} to assess whether the synthesized program captures the intended edit logic.
SSR is defined as the proportion of tasks for which the synthesized program produces hull selections identical to those of the reference DSL program across all non-demonstration meshes.
This behavioral comparison accommodates semantically equivalent but syntactically distinct programs.
A task is counted as successful only if all five developers synthesize programs that satisfy this condition; otherwise, it is counted as failed.
SSR evaluates synthesis quality within MeshForge's DSL, but does not assess whether the edited meshes achieve the intended interaction behavior.
Therefore, we report \textit{Functional Pass Rate (FPR)}, the proportion of non-demonstration meshes whose edited collision mesh satisfies a predefined functional criterion.
Following the functional evaluation in prior work~\cite{wei2022approximate,andrews2024navigation}, criteria are defined per task based on the intended interaction behavior and verified using one of three methods: (1) \textit{physics simulation} in the Unity game engine~\cite{unity2005}; (2) \textit{automated script checks}; or (3) \textit{rule-based structural layout checks}.
We also report the number of demonstrations (\#Demo) and time as effort metrics.
Continuous measurements are first averaged across developers for each task and then reported as mean\,$\pm$\,standard deviation across tasks.

\begin{table}[t]
	\centering
	\small
	\caption{RQ1 results.
		\textbf{N}: number of tasks.
		\textbf{\#Demo}: demonstrations used per task.
		\textbf{Synth.}: program synthesis time (s).}
	\label{tab:rq1}
	\begin{tabular}{l r c c c c}
		\toprule
		\textbf{Difficulty} & \textbf{N} & \textbf{\#Demo} & \textbf{SSR} & \textbf{FPR}  & \textbf{Synth.\,(s)} \\
		\midrule
		Easy                & 8          & $1.1\pm0.1$     & 8/8          & $1.00\pm0.01$ & $1.2\pm0.6$          \\
		Medium              & 8          & $2.3\pm0.5$     & 8/8          & $0.92\pm0.05$ & $2.2\pm0.9$          \\
		Hard                & 8          & $3.4\pm0.7$     & 7/8          & $0.85\pm0.06$ & $7.0\pm2.1$          \\
		All                 & 24         & $2.2\pm1.1$     & 23/24        & $0.92\pm0.08$ & $3.5\pm2.9$          \\
		\bottomrule
	\end{tabular}
\end{table}

\subsection{RQ1 (Effectiveness)}
\label{sec:rq1}
RQ1 evaluates whether MeshForge can turn a few demonstrations into reusable batch-editing programs that generalize across heterogeneous meshes, at two levels: program synthesis (SSR) and editing results (FPR).
Tab.~\ref{tab:rq1} summarizes the main results.
Overall, MeshForge achieves an SSR of 95.8\% (23/24 tasks), with a mean of 2.2 demonstrations per task and an average synthesis time of 3.5\,s.
The performance degrades with task difficulty: Easy and Medium tasks are all successfully synthesized (8/8 each), while Hard tasks achieve an SSR of 7/8 (87.5\%).
Both the number of demonstrations and synthesis time increase monotonically with difficulty, suggesting that harder tasks require more demonstrations and longer search.

\head{Synthesis Failure.}
The single synthesis failure is a \emph{Hard} task in which three of five developers failed to synthesize the correct program.
This task involves multiple rules, where the failing rule is to replace ground-contact hulls only on floor-standing lamp bases, which requires \op{Geo}(\op{Is}(\semlabel{base}), $-Z$).
Because ceiling-mounted lamps are rare in the dataset, the simpler \op{Is}(\semlabel{base}) is \emph{observationally equivalent} to the correct extractor on the provided demonstrations, and thus some developers accepted it.
This reflects a fundamental PBE challenge: multiple programs may be consistent with the demonstrations yet generalize differently on rare asset variants.

\begin{figure}[t]
	\centering
	\includegraphics[width=\linewidth]{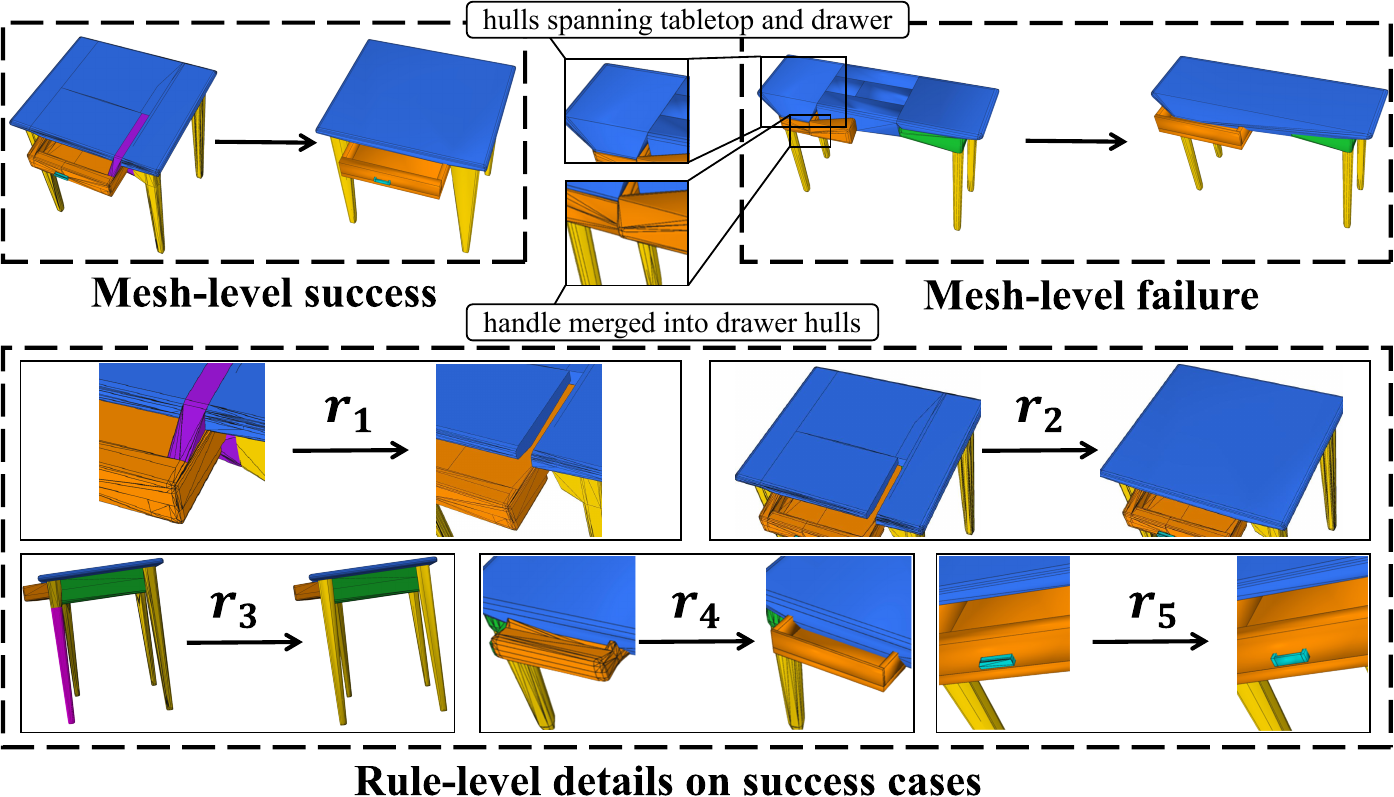}
	\caption{Representative mesh-level success and failure cases (top), and rule-level editing details on success cases (bottom).}
	\label{fig:rq1-func-val}
\end{figure}

\head{Functional Validation.}
Beyond task-level SSR, FPR evaluates the edited collision meshes at the mesh level.
MeshForge achieves an overall FPR of $0.92\pm0.08$, with $1.00\pm0.01$, $0.92\pm0.05$, and $0.85\pm0.06$ for Easy, Medium, and Hard tasks, respectively.
Fig.~\ref{fig:rq1-func-val} shows representative successful and failed final editing results from the table task used as the motivating example in \S\ref{sec:motivating-example} and \S\ref{sec:running-example}.
Among the failed cases, we identify a systematic failure pattern: when the initial convex decomposition algorithm merges task-critical semantic parts into the same hull, hull-level editing cannot recover the missing structure regardless of the number of demonstrations.
For example, in the failed case of Fig.~\ref{fig:rq1-func-val}, several hulls span both the tabletop and drawer region, and the handle is merged into drawer hulls, causing the hull-level edits to fail.
We treat such cases as decomposition failures outside MeshForge's editing scope: the system assumes semantically distinct parts are represented as separate hulls, while recovering the missing structure requires re-decomposition or sub-hull reconstruction.

\head{Few-shot Sensitivity.}
We analyze how SSR evolves as $k$ increases from one to five, using the first successful synthesis step recorded for each developer in the main experiments.
As shown in Fig.~\ref{fig:rq1-few-shot}, all SSR curves are non-decreasing with the number of demonstrations.
The overall curve rises from 33.3\% at \(k=1\) to 95.8\% at \(k=5\), with the steepest improvement between \(k=2\) and \(k=3\).
Easy tasks saturate at \(k=1\) (100.0\% SSR), indicating that a single demonstration suffices for simple tasks.
Medium tasks reach 100.0\% SSR at \(k=3\), while Hard tasks still show one unresolved failure at \(k=5\) (87.5\%), consistent with the synthesis failure above.

\begin{tcolorbox}[rqbox]
	\textbf{Answer to RQ1:} MeshForge achieves 95.8\% SSR (one failure) and $0.92\pm0.08$ FPR, reaching 91.7\% SSR with $k{=}3$ demonstrations, showing strong few-shot efficiency.
\end{tcolorbox}

\begin{table}[t]
	\centering
	\small
	\caption{Qualitative comparison with existing approaches. (\ding{51}: Supported, $\bigcirc$: Partially Supported, $\times$: Not Supported)}
	\label{tab:rq2_qualitative_comparison}
	\begin{tabular}{l c c c c}
		\toprule
		\textbf{Approach}  & \textbf{Legal.} & \textbf{Gen.} & \textbf{Interp.} & \textbf{Input} \\
		\midrule
		Vinedresser3D      & $\times$        & $\times$      & $\times$         & Text           \\
		TextDeformer       & $\times$        & $\times$      & $\times$         & Text           \\
		ParSEL             & $\bigcirc$      & $\times$      & \ding{51}        & Text + labels  \\
		BlenderAlchemy     & $\times$        & $\times$      & \ding{51}        & Text/image     \\
		Manual             & $\bigcirc$      & $\times$      & \ding{51}        & GUI            \\
		\midrule
		\textbf{MeshForge} & \ding{51}       & \ding{51}     & \ding{51}        & GUI demos      \\
		\bottomrule
	\end{tabular}
\end{table}

\subsection{RQ2 (Comparison)}
\subsubsection{Qualitative Comparison}
\label{sec:qualitative-comparison}
To our knowledge, no existing approach is designed for collision mesh batch-editing.
We first compare related approaches qualitatively to clarify why this domain requires a specialized workflow (Tab.~\ref{tab:rq2_qualitative_comparison}).

\head{Physical Legality (Legal.).}
Vinedresser3D~\cite{chi2026vinedresser3d}, TextDeformer~\cite{gao2023textdeformer}, and BlenderAlchemy~\cite{huang2024blenderalchemy} do not enforce convexity on their outputs, which could lead to geometries inadmissible for physics engines.
ParSEL~\cite{ganeshan2024parsel} partially preserves legality under basic affine operations.
Manual editing can preserve physical legality when the developer explicitly maintains convexity.
In comparison, MeshForge operates at the hull level, ensuring convexity of every output hull.

\head{Cross-Instance Generalization (Gen.).}
Existing approaches cannot synthesize reusable editing logic across heterogeneous collision meshes, whereas MeshForge synthesizes a symbolic DSL program that does.

\head{Interpretability (Interp.).}
Vinedresser3D and TextDeformer output only the modified geometry, providing no human-readable record of what was changed.
Manual editing is inherently interpretable, as developers directly perform and can track each change.
ParSEL exposes a parameterized DSL program, BlenderAlchemy outputs executable Python scripts, and MeshForge produces symbolic DSL programs, all providing interpretable artifacts for users to inspect.

\head{User Input.}
Vinedresser3D and TextDeformer accept natural language prompts, while BlenderAlchemy also accepts reference images.
ParSEL additionally requires a pre-segmented, semantically labeled mesh as input.
Manual editing operates through direct per-instance editing.
MeshForge uses GUI demonstrations, allowing users to specify complex edits through direct hull selection and actions.

\begin{figure}[t]
	\centering
	\begin{minipage}[t]{0.48\columnwidth}
		\includegraphics[width=\linewidth]{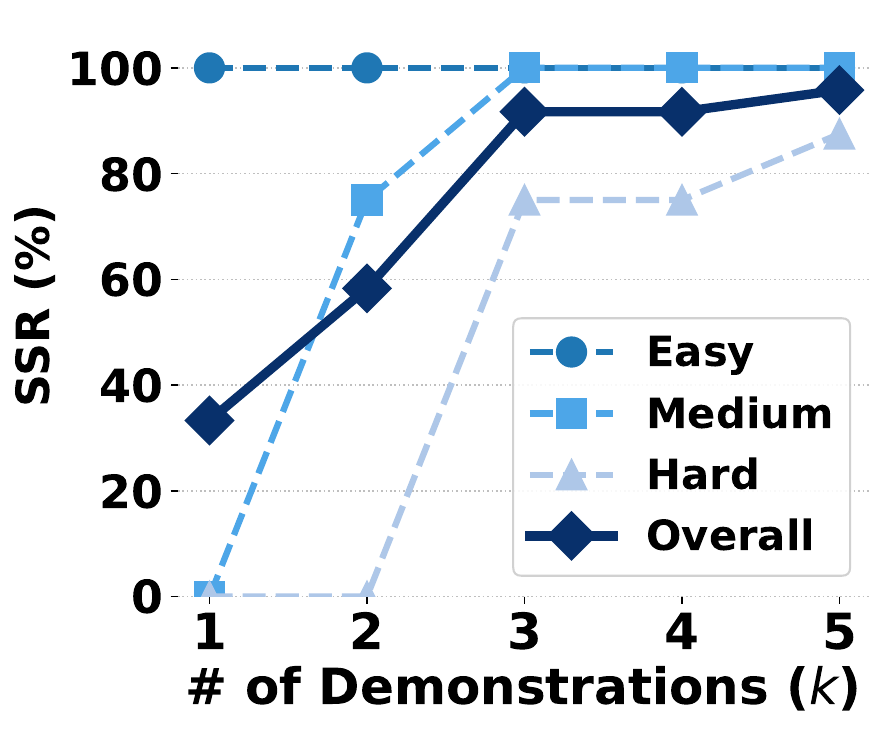}
		\captionof{figure}{Few-shot sensitivity by difficulty level.}
		\label{fig:rq1-few-shot}
	\end{minipage}
	\hfill
	\begin{minipage}[t]{0.48\columnwidth}
		\includegraphics[width=\linewidth]{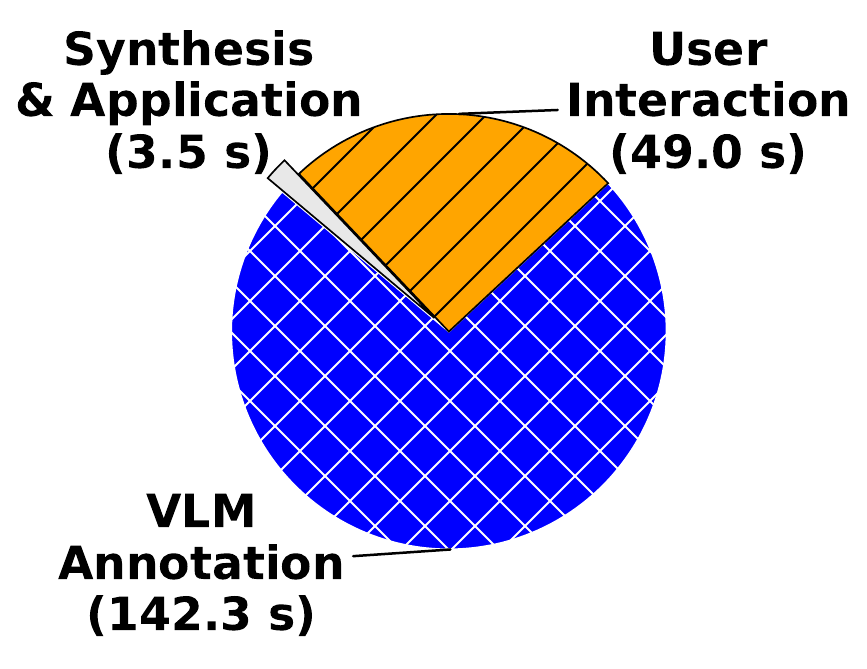}
		\captionof{figure}{Average time breakdown per task by phase.}
		\label{fig:time}
	\end{minipage}
\end{figure}

\subsubsection{Quantitative Comparison}
\label{sec:quantitative-comparison}
Since no prior system targets collision mesh batch-editing, we compare MeshForge with two adapted VLM baselines.
Fig.~\ref{fig:time} shows that VLM annotation and user interaction dominate end-to-end time, with program synthesis accounting for a negligible fraction.
This makes comparisons against alternative synthesis backends (e.g.,~\cite{barbosa2022cvc5,alur2017scaling}) of limited practical value: the bottleneck lies in perception and user interaction, not synthesis.
The more meaningful comparison is whether VLMs can replace the PBE workflow altogether, either by directly generating DSL programs or by scripting the editing environment.
We therefore implement two baselines:
\textbf{VLM Agent.}
Inspired by neuro-symbolic visual programming~\cite{gupta2023visual}, this baseline prompts GPT-4o to synthesize a MeshForge DSL program directly.
GPT-4o receives the DSL grammar, symbolic hull annotations, and rendered views.
It supports iterative refinement, receives the same demonstrations as MeshForge, and is evaluated under the same protocol.
This baseline tests whether direct VLM program generation can replace MeshForge's enumerative synthesis within the same DSL.
\textbf{VLM Blender.}
Adapted from BlenderAlchemy~\cite{huang2024blenderalchemy}, this baseline prompts GPT-4o to generate Blender Python scripts that directly manipulate collision mesh.
It receives the same rendered views and demonstrations, and is free to decide between a batch script and per-instance editing.
This baseline tests whether general-purpose editing scripts can bypass the MeshForge DSL and PBE abstraction.
MeshForge outperforms both baselines (FPR $0.92\pm0.08$ vs.\ $0.69\pm0.12$ for VLM Agent and $0.52\pm0.14$ for VLM Blender).

\begin{tcolorbox}[rqbox]
	\textbf{Answer to RQ2:}
	MeshForge bridges the gap for collision mesh batch-editing, and it outperforms both the VLM Agent and VLM Blender baselines.
\end{tcolorbox}

\subsection{RQ3 (Ablation Study)}
\label{sec:rq3}
RQ3 evaluates two core design choices: the three-layer extractor design and ABI-based label correction.

\begin{figure}[t]
	\centering
	\begin{minipage}[t]{0.48\columnwidth}
		\centering
		\includegraphics[width=\linewidth]{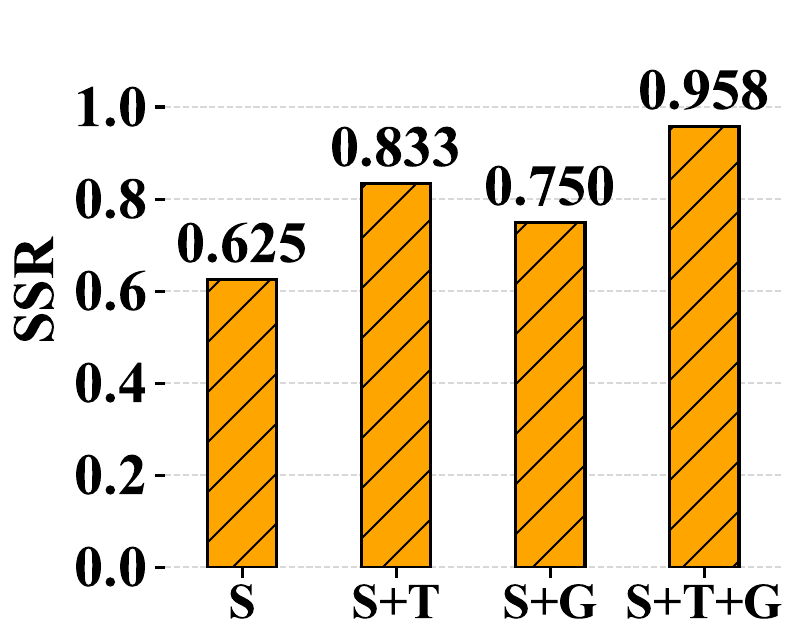}
		\captionof{figure}{Ablation study of the three-layer extractor design.}
		\label{fig:rq3-part1}
	\end{minipage}
	\hfill
	\begin{minipage}[t]{0.48\columnwidth}
		\centering
		\includegraphics[width=\linewidth]{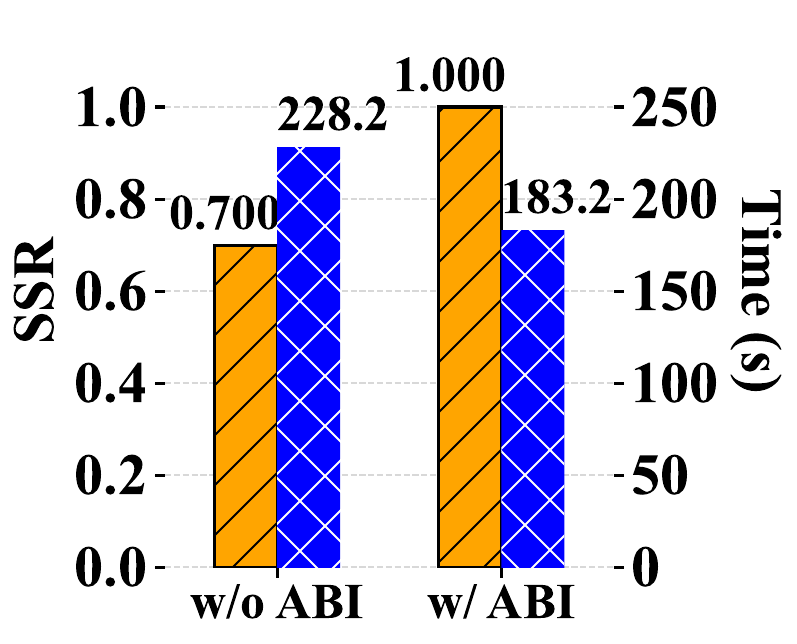}
		\captionof{figure}{Ablation study of ABI-based label correction.}
		\label{fig:rq3-part2}
	\end{minipage}
\end{figure}

\head{Extractor Design.}
We compare the full three-layer design (\textbf{S+T+G}) against three ablated variants:
\begin{enumerate*}[label=(\arabic*)]
	\item semantic-only extraction (\textbf{S}),
	\item semantic and topological extraction (\textbf{S+T}), and
	\item semantic and geometric extraction (\textbf{S+G}).
\end{enumerate*}
This ablation investigates whether semantics, attachment relationships, and spatial extremity each contribute distinct selection power.
As shown in Fig.~\ref{fig:rq3-part1}, removing the geometric layer (\textbf{S+T}) drops SSR to 0.833 (vs.\ 0.958), as the extractor can no longer refine candidates to spatial boundary subsets, such as tabletop support surfaces or ground-contact hulls.
Removing the topological layer (\textbf{S+G}) degrades SSR further to 0.750, as the extractor can no longer distinguish hulls by attachment, such as handles attached to drawers versus other parts.
Semantic-only extraction (\textbf{S}) suffers most (0.625), losing both structural and spatial refinement.
This indicates that all three layers are complementary for accurate hull selection.

\head{Abductive Inference.}
On the main benchmark, ABI was triggered on average in 10.4 of 24 tasks per developer, requiring only 3.6 confirmation decisions per task, far cheaper than manually inspecting every hull label across all demonstration meshes.
A potential concern is that ABI may fit labels to an overly simple, incorrect program.
The user-confirmation step safeguards against this: a correction is applied only if the user inspects and confirms it.
Across all triggered instances in our evaluation, every confirm/reject decision agreed with the label reference (100\% confirmation accuracy), and no accepted correction led to a synthesized program that disagreed with the reference program (zero false-accepts).

To study ABI's effect empirically under a controlled and reproducible setting, we assemble a fixed set of 10 ABI-triggered task instances where all conditions start from the same fixed pool of selected demonstration meshes, so the near-miss condition is identical for every developer and the only manipulated variable is the noise-correction mechanism.
Fixing the demonstrations makes triggering deterministic.
The same five developers complete these tasks, and all conditions use the same precomputed annotations.
Developers practice beforehand, and the order of conditions is counterbalanced to mitigate learning effects.
As shown in Fig.~\ref{fig:rq3-part2}, without any noise correction mechanism (ABI disabled, no manual correction), near-miss programs are unconditionally rejected, dropping SSR from 1.00 to 0.70 and inflating average end-to-end time from 183.2\,s to 228.2\,s.
The inflation stems from additional demonstration rounds when synthesis fails.

We further compare MeshForge against Standard PBE, which disables ABI, so users must manually inspect and correct erroneous labels.
We report SSR and end-to-end time as objective metrics, and ask developers to rate two subjective dimensions on a 1-to-5 scale:
\textit{Cognitive Burden} (``How mentally demanding was the correction?'') and
\textit{Interaction Efficiency} (``How efficiently could you make the correction?'').
As summarized in Tab.~\ref{tab:abi-study}, MeshForge achieves the same SSR while reducing time, and is rated more favorably on both subjective dimensions.

\begin{tcolorbox}[rqbox]
	\textbf{Answer to RQ3:}
	Both the three extractor layers and ABI are essential: the former capture reusable edit intent, while ABI corrects sparse label errors that otherwise derail synthesis.
\end{tcolorbox}

\begin{table}[t]
	\centering
	\small
	\caption{Comparison of MeshForge and Standard PBE. Ratings are 1--5; $\uparrow$ higher is better, $\downarrow$ lower is better.}
	\label{tab:abi-study}
	\begin{tabular}{l c c c c}
		\toprule
		             & \textbf{SSR}
		             & \textbf{Time (s)\,$\downarrow$}
		             & \textbf{\makecell{Cognitive\\Burden\,$\downarrow$}}
		             & \textbf{\makecell{Interaction\\Efficiency\,$\uparrow$}}                     \\
		\midrule
		Standard PBE & 10/10                                                   & 209.6 & 4.2 & 2.6 \\
		MeshForge    & 10/10                                                   & 183.2 & 1.6 & 3.4 \\
		\bottomrule
	\end{tabular}
\end{table}

\section{Threats to Validity}
\label{sec:threats_to_validity}
\head{Internal Validity.}
First, MeshForge's effectiveness relies on the underlying neural model.
If the neural model fails to reliably recognize a class of hulls, MeshForge may synthesize incorrect programs.
The topological and geometric extractors reduce reliance on labels alone, and the ABI algorithm corrects sparse annotation errors, though they cannot fully eliminate them.
Second, MeshForge's DSL does not support more complex editing operations, such as conditional logic, proportion-based geometric constraints, and vertex-level editing.
We will open-source MeshForge to enable community extensions.
Third, MeshForge is human-in-the-loop: users demonstrate edits in a specific order and inspect results to decide whether more demonstrations are needed.
Such human judgment is subjective and may introduce variability.
More systematic demonstration selection~\cite{ji2020question} is a potential mitigation.

\head{External Validity.}
First, we evaluate on a self-constructed benchmark, as MeshForge is the first work to address this problem.
Its scale and diversity (asset categories, sources, and decomposition strategies) are limited.
Future work will validate MeshForge on a larger and more diverse benchmark and conduct larger-scale user studies to quantify user experience.
Second, MeshForge's neural component relies on the GPT-4o API, which introduces API costs, network latency, and data privacy risks.
Replacing it with a local open-source model is a potential mitigation, though open-source VLMs underperform on hull annotation, a direction we leave to future work.

\section{Related Work}
\head{3D Asset Maintenance.}
Existing mesh editing approaches include text-driven geometry deformation~\cite{gao2023textdeformer}, agentic 3D editing~\cite{chi2026vinedresser3d}, parametric shape editing~\cite{ganeshan2024parsel}, and editing script generation~\cite{huang2024blenderalchemy}.
These systems primarily target visual or geometric asset editing.
Andrews~\cite{andrews2024navigation} instead addresses a collision mesh challenge: convex decompositions that obstruct navigable openings.
PhysX-3D~\cite{cao2026physx} and PhysForge~\cite{yang2026physforge} highlight the importance of physics-grounded 3D assets in interactive 3D software, but focus on physical properties such as material and kinematic constraints rather than collision meshes.
Unlike these approaches, MeshForge targets collision mesh batch-editing: it preserves the convexity constraints required by physics engines, synthesizes reusable batch-editing programs, and addresses multiple functional editing motifs.

\head{Program Synthesis.}
Program synthesis supports a broad range of SE tasks, including code transformation~\cite{jiang2019inferring}, test generation~\cite{park2021jest}, and program repair~\cite{nguyen2013semfix,mechtaev2016angelix,xiong2017precise}.
Recent work integrates large language models (LLMs) with program synthesis: Jigsaw~\cite{jain2022jigsaw} augments LLM-generated code with program analysis to ensure semantic correctness, while PyCraft~\cite{dilhara2024unprecedented} combines LLMs with example-driven synthesis to automate code changes at scale.
PBE, an inductive form of program synthesis, has been widely applied in SE.
LASE~\cite{meng2013lase} applies PBE to systematic code transformation.
Rolim et al.~\cite{rolim2017learning} synthesize syntactic program transformations, enabling systematic batch code changes.
Beyond code, PBE has been applied to image processing~\cite{barnaby2023imageeye}, mobile app synthesis~\cite{li2022push}, and data transformation~\cite{wu2023programming}.
Prior work also combines neural perception with symbolic synthesis (neuro-symbolic PBE)~\cite{barnaby2023imageeye,yi2018neural}.
MeshForge extends this line from image, app, and data artifacts to symbolic collision meshes in 3D software.

\head{3D Software Engineering.}
The recent proliferation of 3D software has motivated SE research in this domain.
Li et al.~\cite{li2026} propose a constraint-expressive IR to guide the synthesis of 3D software.
GameRTS~\cite{yu2023gamerts} introduces regression test selection for 3D game software.
Ren et al.~\cite{ren2025reinforcement} conduct reinforcement learning-based fuzz testing on robotic simulator software.
Studies of VR applications have further uncovered diverse reliability issues: Li et al.~\cite{li2024less} identify stereoscopic visual inconsistencies that can lead to cybersickness in VR applications, while Guo et al.~\cite{guo2024empirical} conduct an empirical analysis of security and privacy vulnerabilities.
Zuo et al.~\cite{zuo2023peek} detect 3D model clones across mobile games, revealing the extent to which 3D assets are reused and managed in production pipelines.
Additionally, the complexity of 3D scenes and assets introduces unique challenges in automated 3D software testing~\cite{zhu2025vrexplorer,wang2022vrtest,wang2023vrguide,li2026fse,rzig2023virtual}.
Collectively, these works demonstrate that 3D software presents diverse challenges encompassing reliability, security, and asset management.
MeshForge addresses a complementary asset-management challenge: the labor-intensive batch-editing of collision mesh assets.

\section{Conclusion}
We present MeshForge, which bridges a research gap in 3D software: synthesizing batch-editing programs for collision meshes, a software-artifact maintenance task. 
More broadly, like the batch transformation and refactoring of source code, this task can be automated through program synthesis instead of repeated manual effort. 
We hope MeshForge inspires further research in 3D software engineering.

\bibliographystyle{IEEEtran}
\bibliography{myreference}

\end{document}